\documentclass[conference]{IEEEtran}
\IEEEoverridecommandlockouts
\usepackage{comment}
\usepackage{threeparttable}
\usepackage{listings,lstautogobble}
\usepackage{enumitem}
\usepackage{multirow}
\usepackage{array}
\newcolumntype{L}[1]{>{\raggedright\let\newline\\\arraybackslash\hspace{0pt}}m{#1}}
\usepackage{cite}
\usepackage{amsmath,amssymb,amsfonts}
\usepackage{algorithmic}
\usepackage{graphicx}
\usepackage{textcomp}
\usepackage{xcolor}
\def\BibTeX{{\rm B\kern-.05em{\sc i\kern-.025em b}\kern-.08em
    T\kern-.1667em\lower.7ex\hbox{E}\kern-.125emX}}

\usepackage{makecell}
\usepackage{csquotes}
\usepackage[none]{hyphenat}
\begin{document}

\title{FLASH 1.0: A Software Framework for Rapid Parallel Deployment and Enhancing Host Code Portability in Heterogeneous Computing}

\author{Michael Riera, Masudul Hassan Quraishi,
        Erfan Bank Tavakoli,
        Fengbo Ren\\
        }

\maketitle

\begin{abstract}
This paper presents FLASH 1.0, a C++-based software framework for rapid parallel deployment and enhancing host code portability in heterogeneous computing. FLASH takes a novel approach in describing kernels and dynamically dispatching them in a hardware-agnostic manner. FLASH features truly hardware-agnostic frontend interfaces, which unify the compile-time control flow and enforce a portability-optimized code organization that imposes a demarcation between computational (performance-critical) and functional (non-performance-critical) codes as well as the separation of hardware-specific and hardware-agnostic codes in the host application.  We use static code analysis to measure the hardware independence ratio of twelve popular HPC applications and show that up to 99.72\% code portability can be achieved with FLASH. Similarly, we measure and compare the complexity of state-of-the-art portable programming models to show that FLASH can achieve a code reduction of up to 4.0x for two common HPC kernels while maintaining 100\% code portability with a normalized framework overhead between 1\% - 13\% of the total kernel runtime. The codes are available at [hidden-for-double-blind-review].
\end{abstract}

\begin{IEEEkeywords}
High Performance Computing (HPC), heterogeneous computing, Portability, CUDA, OpenCL
\end{IEEEkeywords}

\section{Introduction}

\begin{figure*}[!]
  \centering
  \includegraphics[width=7in]{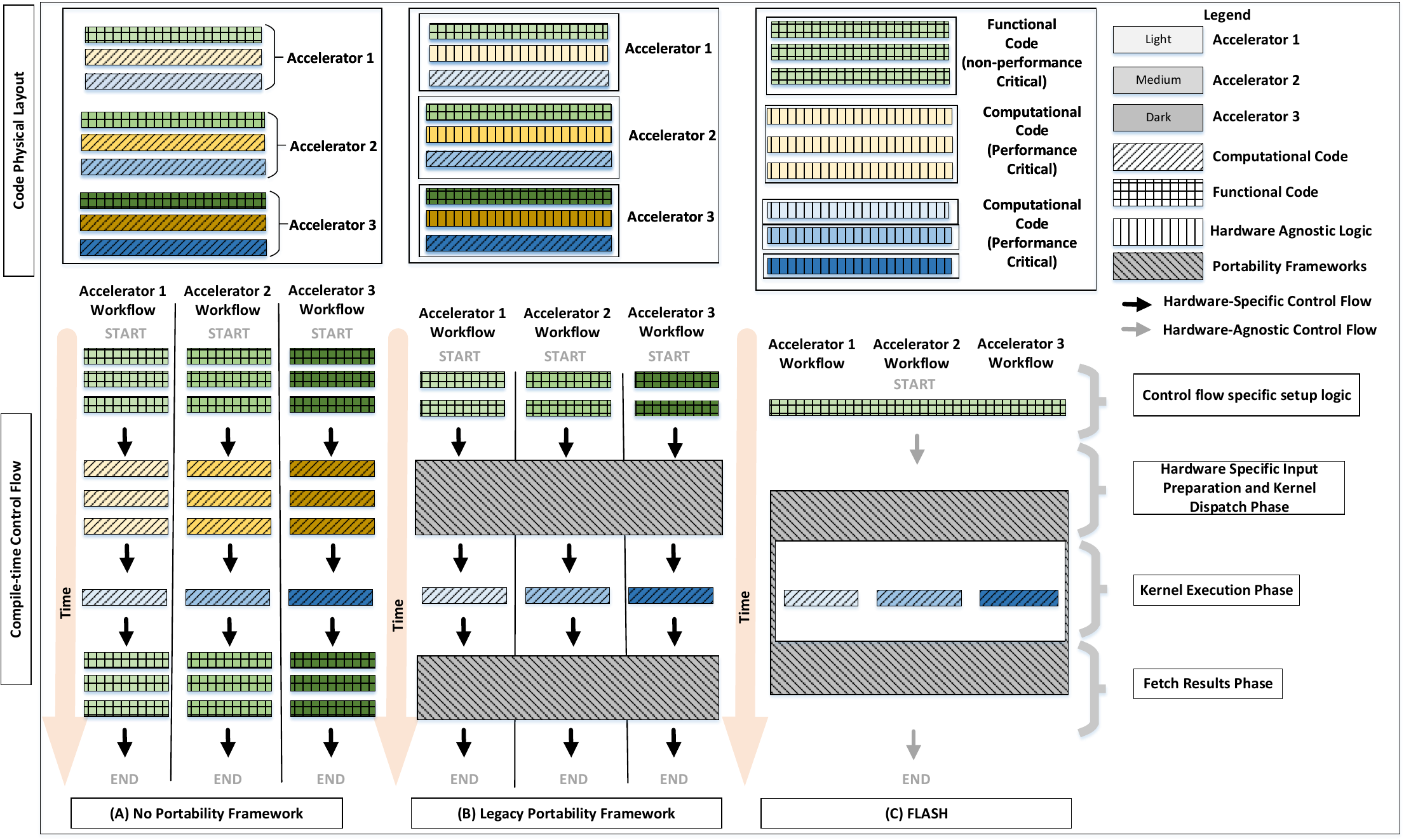}
  \caption{Illustration of the key differences in the pattern of code placement and compile-time control flow among: a) no portability framework, b) existing portability frameworks, and c) FLASH 1.0. The hardware-agnostic interfaces, the unification of flow control, and the separation of functional and computational software with line of source code reduction are the key to fast parallel deployment and portability enhancement.}
  \label{Figure: EXISTING_WORK_COMP}
\end{figure*}

As HPC software becomes more sophisticated and complex (\textit{i.e.}, multi-disciplinary and multi-platform), frameworks and programming models providing host code portability guarantees are becoming increasingly critical for minimizing the overhead of refactoring application codes to maintain correctness and cutting-edge performance across various hardware platforms \cite{8049026} \cite{10.1145/3318170.3318196} \cite{10.1145/3380536.3380542} \cite{10.1145/3318170.3318194} \cite{10.1145/3388333.3388653}. 
As more accelerator platforms become increasingly domain-specific and widely available, HPC developers are faced with the problem of perpetual refactoring of application codes in order to continue making ground-breaking science practical. Code refactoring often requires a large amount of time and expertise of HPC developers to understand the new accelerator hardware, runtime interfaces, and application architecture. As a result, frameworks for improving code portability have taken center stage as a means to reduce the amount of code refactoring effort necessary to facilitate the adoption of new hardware accelerators in HPC. Moreover, with the performance of general-purpose computing quickly plateauing, HPC software solutions must look toward more application- and domain-specific accelerators (ASAs/DSAs) to reach the next notable milestones in performance \cite{10.1145/3332186.3332223} \cite{10.1145/3426428.3426915}. This will undoubtedly increase the cadence of code refactoring to an impractical level with existing solutions and, more so, without. 

The existing solutions (\textit{e.g.}, SyCL variances, HIP, Legion, and others) for improving portability in HPC fall notably short in providing user-friendly and truly hardware-agnostic functionality to the overlaying application \cite{10.1145/3204919.3204939}, \cite{10.1145/3388333.3388641}, \cite{10.1145/3388333.3388643}. Existing solutions only provide complex, pseudo-hardware-agnostic interfaces and hardware-specific primitives that are predominately compatible with general-purpose or  Von Neumann architecture-based platforms. Furthermore, these solutions only offer minimal portability guarantees as they fail to decouple framework logic (non-performance-critical) components from the computational (performance-critical) kernels in a systemic way, nor do they unify the host code control flow (see Figure. \ref{Figure: EXISTING_WORK_COMP}). Such separation is critical for reducing the code refactoring efforts toward adopting new accelerator hardware. Due to the lack of explicit code organization, the existing solutions rely on HPC developers to further reduce the complexity of porting applications by means of good coding practices, which adds undue burden to both the development and maintenance of HPC applications in general. Lastly, the focus of existing portability solutions attempts to envelop both host and kernel code ignoring the effectiveness and the performance benefits of using native libraries, languages, and runtimes. We offer a different perspective towards portability that focuses on working alongside native components for performance while maximizing portability via a robust active host-side API.   

  \begin{figure}[!]
  \centering
  \includegraphics[width=2.5in]{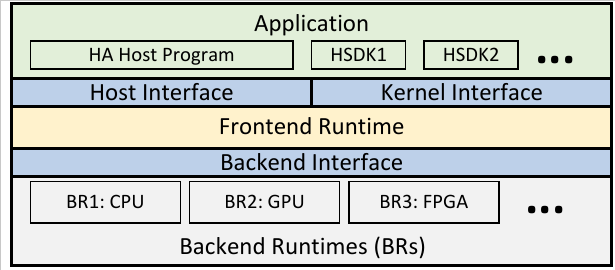}
  \caption{Block diagram of FLASH 1.0.}
  \label{Figure: FLASH_BLOCK_ARCH}
\end{figure}

This paper presents FLASH 1.0, a software framework for rapid parallel deployment and enhancing host code portability in heterogeneous computing. FLASH 1.0 is a C++-based framework that critically serves as a clear way-point for separating hardware-agnostic and hardware-specific logic to facilitate code portability. FLASH 1.0 uses \textit{variadic} templates and \textit{mixin} idioms-based interfaces as the primary vehicle to enable simple, extensible hardware-agnostic interfaces easily supportable by legacy and future accelerators of various architectures. FLASH 1.0 consists of four major components (see Figure \ref{Figure: FLASH_BLOCK_ARCH}): the frontend (host and kernel) and backend interfaces and the frontend and backend runtimes.

The frontend interfaces provide facilities to engage, dispatch, and build deep kernel pipelines and strongly decouple the host program from kernel implementations. Such an interface strategy provides the application with extensible hardware-agnostic interfaces that enforce a clear separation of responsibility among hardware-agnostic host applications, hardware-specific device kernels, and runtimes. The variadic and mixin idioms allow for a much smaller set of interfaces that covers a larger region of functionality with little repetition as compared to existing portability frameworks. Furthermore, with the simple and loosely-coupled natures of the frontend interfaces, developers can easily organize, distinguish, maintain, and extend hardware-specific device kernel (HSDK) definitions and accelerator runtime interfaces without disturbing the existing logic. This accelerates the adoption of new, exotic accelerators by no longer requiring optimization or hardware engineers to understand the application as a whole to enable new functionality.    

The frontend runtime packages arguments, checks dependencies, translates attributes of complex kernel pipelines and dispatches them to an appropriate backend runtime. Moreover, the frontend runtime has the novel responsibility for converting an open-ended meta-programming interface semantic to a traditional close-ended \textit{PImpl} interface for backend modularity. This is critical to providing practical extensibility to an application without increasing compilation time to an impractical level. Moreover, with a compiler-firewalled frontend runtime, applications, and device runtimes can evolve independently, which reduces development time by allowing application developers and hardware engineers to work independently.

The backend interface is a close-ended interface specification that allows for custom backends to be added independently of all other components in the Flash framework. The backend runtime wraps and bridges the respective device runtimes to a frontend-compatible interface. The backend device runtimes provide encapsulation for hardware-specific functionality (\textit{e.g.}, API or runtime ) that is compiler-firewalled from other components in the FLASH framework for independent development. Device runtimes must adhere to the backend interface and self-register. This strategy allows for device runtimes to be developed independently and quickly without disturbing or requiring familiarity with the overall application or knowledge of adjacent accelerators. 

The contributions of this work are summarized as follows:
\begin{itemize}[itemsep=0pt,topsep=0pt]
\item We propose FLASH 1.0, a C++-based software framework for rapid parallel deployment and enhancing host code portability in heterogeneous computing. 
\item We propose truly hardware-agnostic frontend interfaces, which not only unify the compile-time control flow but also enforce a portability-optimized code organization that imposes a clear demarcation between computational (performance-critical) and functional (non-performance-critical) codes as well as the separation of hardware-specific and hardware-agnostic codes in the host application.
\item We propose a new kernel deployment model for transparently scaling up kernel execution across multiple heterogeneous accelerators on a single node with built-in automated synchronization and partitioning.
\item We develop the device runtimes for supporting Intel CPUs, NVIDIA GPUs, and Intel FPGA devices in FLASH.
\item We analyzed the complexity of different portability frameworks, including DPC++, KOKKOS, HIP, RAJA, Legion, and FLASH by measuring the number of interfaces (including methods and objects) available to the user. The analysis shows that FLASH 1.0 is 13\% - 90\% - less complex than the aforementioned frameworks in terms of host code interfaces. 
\item We quantitatively define framework portability and analyze the framework portability of the existing solutions and FLASH. The analysis shows that the portability of the existing solutions ranges from 23\%-91\%, whereas FLASH 1.0 provides a 100\% portability guarantee.
\item We compare source line of codes (SLOC) in terms of host-side API calls (\textit{e.g.} \textit{\#pragmas, parallel\_for}) or API object construction (\textit{e.g.} \textit{cpu\_selector, RuntimeObj}), framework overhead and kernel runtime of two benchmark host codes (N-Body and Particle Diffusion) implemented using four existing solutions (KOKKOS, RAJA, OpenMP, SyCL) and FLASH on CPU and GPU. We observe that FLASH 1.0 is able to achieve a 1-2.2$\times$ reduction in the source line of codes (SLOC). The reduction mainly comes from initialization, data preparation, and synchronization.
\item We compared the normalized framework overhead of two kernels, N-Body and Particle Diffusion and showed that FLASH takes between 1\% - 13\% overhead of the total runtime, with OpenMP inducing as much as 24\% overhead on the runtime.
\item We quantitatively defined application code portability and statically analyzed the codebases of 12 common HPC applications to measure their dependency on CUDA throughout the applications at a transnational unit granularity. The analysis projects that FLASH 1.0 can achieve up to 99.72\% of application code portability. 
\end{itemize}

\section{Related Work}
We believe the following four components are key to providing true portability:
\begin{itemize}[itemsep=0pt,topsep=0pt]
\item True hardware-agnostic interface: the same frontend interface can be used for multiple types of accelerators without any modification. 
\item Realtime interoperability and scheduling between accelerators.
\item Predication and loop construction.
\item Fusion, composability, and synchronization.
\item An implied code organization: separation of responsibilities among hardware-agnostic logic, hardware-specific logic, and hardware runtimes.
\end{itemize}

Frameworks such as KOKKOS \cite{10.1145/3204919.3204939}, RAJA \cite{8945721}, DPC++ \cite{10.1145/3388333.3388653}, OpenACC, OpenMP \cite{10.1145/3110355.3110356}, Legion \cite{8916294}, Boost Compute\cite{10.1145/2909437.2909454}, Grid Tools\cite{10.1145/3324989.3325723}, PaRSEC\cite{Parsec}, IRIS\cite{osti_1832700}, and StarPU\cite{AugThiNamWac10RR7240}  are the current state of the art for providing a consistent API to heterogeneous accelerator resources. However, these current solutions fall notably short in providing the mentioned components required to enable a high degree of portability to an application. These solutions share a common theme for dispatching to different backends using tag dispatch. Furthermore, they augment the parallel ‘for’ construct for portability. This is where the lack of portability is centered.

Firstly, current solutions have invalidating coupling between the hardware-agnostic interfaces and kernel execution (at compile-time) that is often represented by hardware-specific template parameters or an input parameter that is either a function pointer, lambda expression, or a task-specific interface (\textit{e.g.}, \textit{cblas\_gemm}, \textit{cudaMemcpy}). Additionally, these strongly-coupled interfaces, such as SyCL's \textit{q.submit} method, or the OpenMP \textit{\#pragma omp...} lead to not only portability issues as they require modification to source code or a binary redeployment re-targeting a different backend with \textit{-fopenmp-target} compiler flags for each supported accelerator but also multiple code paths for each accelerator (and accelerator combination). Furthermore, these solutions have an overwhelming set of interfaces (over 600+), all of which cause alarm when being ported to various hardware platforms for backward compatibility. FLASH alleviates this issue through a combination of reflection, virtual dispatching, and variadic templates. These three idioms/patterns allow the creation of simple, task-centric, extensible, hardware-agnostic interfaces.  

Secondly, current solutions have complex hardware-specific, explicit synchronization schemes for synchronizing groups of hardware-specific processing elements (PEs), requiring either fixed data structures or fine-grain synchronization methods. (\textit{e.g.}, \textit{Kokkos :: Cuda :: fence}, \textit{\_\_threadfence}, or \textit{sycl :: nd\_item.barrier}). Other pragma-based solutions have the added detriment of never being dynamically interoperable or portable without the express recompilation using an associated toolchain for new accelerators. This will be an issue porting to accelerators that do not have a sophisticated toolchain (compiler) to convert OpenMP pragmas to runtime calls. Finally, the generality of these interfaces towards a Von Neumann architecture will make it difficult to port to a non-thread-based accelerator platform (\textit{e.g.}, deep learning accelerator). 

Lastly, existing portability frameworks do not enforce a separation policy between hardware-specific and agnostic code on the overlaying application. This is critical to facilitate the overall portability of the application. In order to further reduce the complexity of refactoring an existing code base to enable emerging hardware accelerators, a clear separation between hardware-agnostic, hardware-specific, and runtime logic is needed. This will make it much easier for developers to work in parallel during the porting activity. For instance, scientists, optimization engineers, and software engineers can work in the hardware-agnostic, hardware-specific, and runtime space, respectively. FLASH uses \textit{pimpl}, \textit{chain-of-responsibility}, \textit{recursive type composition}, \textit{builder}, and \textit{singletons} software idioms/patterns to facilitate code organization, simplify, and maximize functional coverage of the interface, and portability of the application.

\section{FLASH Architecture}

\subsection{Overview}
FLASH is a C++ code organization framework for enhancing host code portability in applications running on heterogeneous computing systems (see Figure \ref{Figure: FLASH_BLOCK_ARCH}) by decoupling an application's performance-critical and non-performance-critical components and unifying the control flow of the application. FLASH makes use of well-known software patterns and idioms (\textit{Builder}, \textit{pImpl}) and new C++ 2020 features and concepts to create simple and intuitive interfaces that maximize functional coverage and portability. This framework serves as a way-point between applications, hardware-specific kernels, and backend runtimes. The FLASH interfaces implement a virtual dispatching and partial reflection capability to invoke and deploy loosely coupled kernels from a single- and/or multi-source heterogeneous compilation process.   

The FLASH framework leverages meta-programming to describe kernel capabilities such as: arguments, kernel traits, parallelism, synchronizations, and partitioning \cite{abrahams2004c++}. The \textit{runtime} objects implement a design pattern named builder pattern to build kernel pipelines and evaluate them on-demand with \textit{exec} interface or \textit{defer} execution with the \textit{defer} interface for lazy evaluation. Runtime objects are non-owning, copyable, and movable. The FLASH interface is completely described by the host interface (including two runtime objects: \textit{RuntimeObj} and \textit{SubmitObj}, and compile-time kernel descriptions) and a FLASH runtime that encapsulates device runtimes. These objects implement a parameter pack tag dispatch interface \cite{TAG_DISPATCH} that can be used to fit any kernel declarations, kernel traits (attributes), indexing schemes (ex. \textit{count\_by}), and containers (such as \textit{std::vector}, or \textit{flash\_memory}).

The FLASH frontend runtime packages arguments, checks dependency, translates kernel traits of complex kernel pipelines and dispatches them to the FLASH backend runtime. The FLASH frontend runtime engine translates the compile-time kernel attributes and argument types to runtime variables and is type-erased and forwarded to the backend runtime. The frontend runtime engine performs deferred indirect type-checking by reconciling the input types of the parameter pack with the types described from the kernel description argument. The FLASH runtime engine parses the kernel pipeline and dispatches kernels to the backend runtime using a round-robin strategy. It also handles the conversion between native index space and FLASH subspace, drives the custom indexing system, and passes it to the backend runtime. FLASH 1.0 supports three widely used device runtimes: CPU, GPU, and Intel FPGA, via jthreads/C++ parallelism, CUDA, and OpenCL, respectively. Finally, FLASH inherits the kernel toolchain-specific interfaces supported by the FLASH backends, respectively, C++ for CPU, CUDA for NVIDIA GPUs, and OpenCL for Intel FPGAs. 

\subsection{Host Interface}

\subsubsection{Kernel Definition Interface}

The semantics for describing kernels rely on two main language features of C++: parameter packs and constexpr constructions. FLASH uses these features to describe kernel argument types, traits, meta-data, synchronization, and partitioning strategies. FLASH's kernel description semantics can be used as a tag to dispatch kernels at runtime. The key contribution driving the development of a meta-programming-based interface is simplicity. Constructing a structured and verbose kernel type with the ability to intuitively reason about its context, parallelism, and synchronization statically will make codes much more human-readable and self-documenting, as well as machine-readable and optimizable. The FLASH kernel description semantics and object define a single fundamental type from which various classes can be derived. The kernel description object encapsulates the associated attribute (kernel-traits) objects that can be used to augment the aforementioned metadata to a kernel definition. 
The "... attrs\_args" (a common C++ vernacular representing a set of data types in metaprogramming) represents the parameter pack that encapsulates various labels, attribute objects, traits, and argument types. 

The kernel description object has required input attribute literals and objects followed by optional attributes and ends with argument declarations. The required section includes the number of pure input arguments and kernel label (\textit{KernelDECL}) to support virtual dispatching. The arguments require strict ordering of read-only and read-write arguments within the kernel description structure. The write-only arguments are dynamic and are evaluated at the call site completing the invocation of a kernel. 

\subsubsection{Memory Buffers}

The FLASH memory buffers are objects that can be used either in the kernel definition or dynamically in the runtime objects. They can either be owner or non-owning data objects. FLASH (memory) buffers have the boiler-plate overloaded operators such as \textit{operator[]}, \textit{at}, \textit{erase} and \textit{size} to query state and access elements therein. By default, the FLASH buffers work as a host or a device buffer and lazily return data at the user-defined byte size at a 4KB page granularity. When the FLASH buffer owns the data, it internally carries a cache footprint of the device data until it explicitly receives a request to move the data to the host (at page boundaries). The kernel definition can have a primitive buffer type \textit{T} while passing a \textit{flash\_memory} container of type \textit{T} at runtime, and the framework will perform the proper conversions. The purpose of this structure is to create a generic interface for device-side buffer manipulation. Additionally, FLASH buffers allow the framework to optimize the data movement when paired with the partitioning scheme, as well as processing kernel pipelines.

\subsubsection{Flash Variable}

The \textit{flash\_variable} has three dimensions: it is used to predicate submissions, loop controls, and to interface with kernel and runtime-side. The predicates are asserted in two ways: one within the \textit{KernelDefinition} and/or within the submission arguments. This gives flexibility for predication to occur with values on the host or device memory (Table \ref{table: PRED_TABLE}) ). Similarly, the loop controls can come from the host, device memory, or kernel/runtime querying. Finally, as an interface to extract values from the runtime and kernel sets. This object is a unified method to interoperate between multiple backends, runtime, and host code.

\begin{figure*}[!]
  \centering
  \includegraphics[width=7.2in]{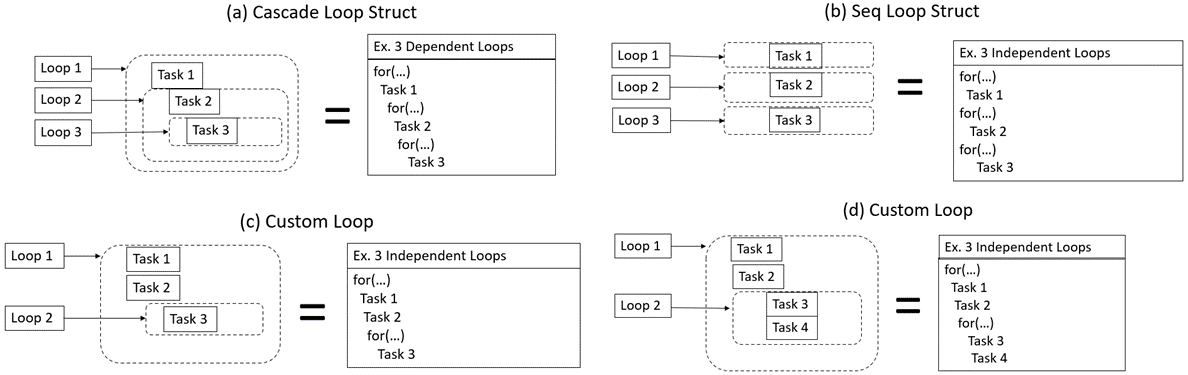}
  \caption{Examples of loop control. (a) : default loop cascading (dependent); (b) : sequential (independent) loop creation; (c) , (d) : custom combinations of dependent and independent loop semantics.}
  \label{Figure: LOOP_STRUCT}
\end{figure*}

\begin{table}[]
\tiny
\lstset{linewidth=8.5cm}
\caption{\label{table: CFD_TABLE} CFD snippet example written in CUDA (top), and rewritten in FLASH (bottom) demonstrating loop semantics. }
\begin{tabular}{|L{8.5cm}|}
\Xhline{0.2ex}
{\textbf{CUDA SLOC: 54 }} \\ \Xhline{0.2ex}
\begin{lstlisting}[belowskip=-0.8 \baselineskip]
  // Begin iterations
  for(int n = 0; n < iterations; n++)
  {
     //move data to device
     //... 
     //for the first iteration we compute the time step
     compute_step_factor<<<gridDim2, BLOCK_SIZE_2>>>(...);

     for(int j = 0; j < RK; j++)
     {
       compute_flux<<<gridDim3, BLOCK_SIZE_3>>>(...);
       time_step<<<gridDim4, BLOCK_SIZE_4>>>(...);
     }
  }
  
\end{lstlisting}  \\ \Xhline{0.2ex}
\end{tabular}
\begin{tabular}{|L{8.5cm}|}
{\textbf{FLASH (hardware-agnostic without recompilation). SLOC: 5 }} \\ \Xhline{0.2ex}
\begin{lstlisting}[belowskip=-0.8 \baselineskip]
using COMP_SF=KernelDefinition<3,"compute_step_factor",...>;
using COMPUTE_FLUX=KernelDefinition<9,"compute_flux",...>;                  
using TIME_STEP=KernelDefinition<5,"time_step",...>;
                                             
void main()
{
  //initialization
  //...
  int iter = 8;
  int RK   = 3;
  auto lv = loop_var(2);
  ocrt.submit(COMP_SF{}, ...)
	    .defer(...)
	  .submit(COMPUTE_FLUX{}, ...)
		.defer(...)
	  .submit(TIME_STEP{}, lv, ...)
		.defer(...)
	  .exec( iters, RK ); //default cascade_loop  
}

\end{lstlisting}  \\ \Xhline{0.2ex}
\end{tabular}
\end{table}

\begin{table}[]
\tiny
\lstset{linewidth=8.5cm}
\caption{\label{table: PRED_TABLE} Example of predicate logic from host and device. }
\begin{tabular}{|L{8.5cm}|}
\Xhline{0.2ex}
{\textbf{Generic Pattern}}  \\ \Xhline{0.2ex}
\begin{lstlisting}[belowskip=-0.8 \baselineskip]
main()
{
  //independent host-side pred logic
  auto pred = Task1(...) //host- or device- logic
  if( pred ){ 
    Task2(...)
  }
  else{ 
    Task3(...) 
  }
}

\end{lstlisting}  \\ \Xhline{0.2ex}
\end{tabular}
\begin{tabular}{|L{8.5cm}|}
{\textbf{FLASH }} \\ \Xhline{0.2ex}
\begin{lstlisting}[belowskip=-0.8 \baselineskip]
using TASK1 = KernelDefinition<...>;
using TASK2 = KernelDefinition<..., pred, ...>;
using TASK3 = KernelDefinition<...>;

template<typename T = bool>
using b_pred_var    = flash_variable<PRED, T>;  

void main()
{
  //initialization
  //...
  auto pred1  = b_pred_var(false);
  auto pred2  = b_pred_var(true); 
  //output predicate from TASK1
  ocrt.submit(TASK1{}, ..., pred1) 
	.defer(...)
 //input predicate from TASK1
      .submit(TASK2{}, pred1, ...) 
	.defer(...)
      .submit(TASK3{}, !pred1 && pred2, ...)
	.exec(...)
}

\end{lstlisting}  \\ \Xhline{0.2ex}
\end{tabular}
\end{table}

\subsubsection{Runtime Object}

The runtime object (\textit{RuntimeObj}) is the object that is used to configure and manipulate both front and backend runtime contexts and kernel pipelines. The constructor is used for backend selection, early registration of kernels with long load times (\textit{e.g.}, FPGA kernels), and associating kernel definitions with implementations. The \textit{RuntimeObj} has only three methods: \textit{options}, \textit{submit}, and \textit{Exec}. The limiting number of interface methods and structures are critical in minimizing the complexity of the framework. The overarching pattern is a builder with transformations to the \textit{SubmitObj}. The \textit{options} method is a general theme in the FLASH framework representing a mixin function, where meta-data can be added to the runtime contexts such as device selection, predicates, and scheduling strategies via object attributes. Additionally, the options method returns a reference to the \textit{RuntimeObj}, allowing for recursive invocations of the options method. Next, the \textit{submit} method generates a submission object (\textit{SubmitObj}). The \textit{submit} method stages a kernel to be launched; it uses tag dispatching and lazy type-checking between the parameter pack and the kernel definition and attaches the arguments to the kernel. 
The submission function can take three additional structs for data-movement optimizations, predication, and kernel/runtime configuration querying. These are the \textit{pred} (for predication), \textit{flash\_variable}, and \textit{flash\_mem} for framework-level memory management and caching. The \textit{pred} allows for a submission to be skipped or executed. The \textit{pred} allocation can reside on the device or host and can also be deferred to the backend runtime (or kernel sandboxes) for a more custom predicate (Table \ref{table: PRED_TABLE}). As discussed in prior sections, the \textit{flash\_mem} and \textit{flash\_variable} are generated and managed by the runtime.

Finally, the \textit{exec} method (from the \textit{RuntimeObj}) controls the loop structure of the transaction. One loop structure defines all the loop characteristics, with an additional two derived helper structures for further simplicity. All loops are defined in terms of triples: offset, stride, and value encapsulated by the \textit{custom\_loop} struct. The \textit{exec} function may have multiple instances of \textit{custom\_loop} in order to define multi-level, multi-look loop execution per transaction. The two helper structs are wrappers around the \textit{custom\_loop} with explicit transaction-level loop strategies. These wrappers are \textit{cascade\_loop}, and \textit{seq\_loop}. By default, with a transaction of N subactions, the exec method employs a \textit{cascade\_loop}, with an offset of one, and a stride of N-'i', where 'i' is the per loop value definition (see Figure \ref{Figure: LOOP_STRUCT}). The \textit{seq\_loop} helper function creates loops with a stride of 1 and an offset of 'i' (Figure \ref{Figure: LOOP_STRUCT}).

\subsubsection{Submission Object}
The submission object is the final stage before submitting the kernel to the FLASH runtime for execution. It contains four methods: \textit{options}, \textit{sizes}, \textit{defer}, and \textit{exec}. Similar to the \textit{RuntimeObj}, the \textit{options} method can dynamically add meta-data to override or add attributes to the submission. The submission object allows the developer to override compile-time kernel traits at the call site for runtime customization. The \textit{sizes} method associates buffer sizes to primitive array types such as int*, ulong*, char*, etc. to the kernel arguments. Additionally, the FLASH frontend runtime extrapolates buffer sizes from range-based and iterable arrays internally, bypassing the explicit nature of most related work.  The \textit{options} and \textit{sizes} both implement a builder pattern to be able to interchange and recursively configure submissions accordingly. The \textit{defer} method returns a runtime object for chaining (or pipelining) kernels together. Submission objects can easily be chained through successive calls to submit (from the runtime object) and defer (from the submission object). The \textit{defer} and \textit{exec} methods take in a work-item listing similar to CUDA, OpenCL, and SyCL with the added benefit of manipulating the indexing scheme beyond the default Cartesian product. These schemes are enabled by a combination of \textit{size\_t} and \textit{count\_by} arguments producing custom sequences in an index table. The CPU device runtime has an N-th dimensional depth, but the GPU and FPGA device runtimes only support three-dimensional execution semantics. The \textit{exec} method executes the entire kernel pipelines (one or more submission objects) and keeps blocking until the chain is completely executed. The submission objects are turned over to the FLASH frontend runtime, where they are scheduled for execution in the context of the \textit{runtime} object. In later versions of FLASH, the runtime will be able to recognize whether a kernel can be fused or if the kernel is an entry point to another kernel. This feature is not yet available but is currently being worked on.

Additionally, there are two structures that associate submission objects with loop controls. The \textit{loop\_vars} allows the kernel to access the loop control variable in instances where the loop values are needed as kernel inputs (see '$lv$' in the CFD benchmarks on Table \ref{table: CFD_TABLE}. The variables are managed host-side and are incremented by the CPU. The \textit{loop\_var} accepts a loop index parameter to indicate which loop control value it is forwarding to the kernel.

\subsection{FLASH Frontend Runtime (FR)}

The FR is the way-point among multiple runtimes (or submission) objects managed by the applications and the device runtimes. The internal to the FR includes three distinct architectural choices: type-erasure of kernel arguments, conversion of kernel traits to runtime variables, and a "Pointer to implementation" (or pImpl) pattern with a backend registry to compiler-firewall the FLASH RT (and backends) for extendability. Before entering the core FLASH engine, the interface between the interface objects exists as a thin conversation layer used to type-erase buffer arguments (including scalars) and convert attribute objects into a structured non-templated descriptor. This allows different template types at the application level to collapse to concrete types for faster compilation and easy manipulation and querying at runtime. It also allows for the FR to use polymorphism to enable self-registering backends while keeping the application compile time low. The FR deals with breaking down the kernel pipeline created at the application layer, checking kernel argument types and dependencies, and dispatching accordingly. By default, the FR submits descriptors (with buffer reference) one by one to the device runtime, chosen by the application layer. If there are no dependency issues, the FR does not wait for a submission to complete before submitting the next one; otherwise, the FR handles the synchronization implicitly. The FR handles partitioning by breaking down a kernel's index table facilitated by a user-defined partitioning function that describes the buffer boundaries as a function of work-item vectors. The FR allocates and maps sub-buffers to FLASH subspaces and flags the execution as a (\textit{subaction}), and merges data back into its original buffer lazily. The FR and device runtimes are singleton patterns to facilitate device runtime compatibility and fine-grain control across multiple processes, devices, and tenants. The FR is reentrant and thread-safe. All interface methods can be safely used for parallel execution. Finally, the device backends follow the FLASH Backend interface and register independently with the FR registrar. 

\subsection{FLASH Backend Runtime (BR) and Device Runtimes (DRs)}

\subsubsection{Backend Interface}

The FLASH 1.0 backend interface allows device runtimes to be wrapped in a FLASH-compatible module for availability to FR. The backend interface implements a simple close-ended interface that indirectly allows the applications to register, execute, and wait for kernel status primarily. The backend interfaces also consist of a self-registering mechanism externally linked to a registrar, which each DR can independently link to. This interface permits each DR to advertise its availability to the frontend runtime system. By virtue of reflection, the backend interface registers kernels in a loosely coupled manner, registering the name and implementation separately, indirectly giving the application the ability to virtually dispatch functionality. By default, the \textit{exec} method will use the implementation of the kernel tied to the initial registration but can be overridden during the creation of the submission, as mentioned in the host interface section. This interface strategy enables portable implementation strategies to be implemented dynamically.      

\subsubsection{CPU Runtime}

The CPU runtime is implemented with two major components: device runtime and the kernel-side interface. Since, at the time of this work, there were no work-item-based parallel command processor runtimes that supported partial reflection, polymorphism, and virtual dispatch, the parallel runtime solution for the CPU is built from the ground up. The CPU runtime uses the linker methods to inspect shared objects or binaries via \textit{dlopen} and \textit{dlsym} to search and invoke kernels dynamically. There is logic in the CPU runtime to convert human-readable method labels for free functions, namespaces, or member functions. The CPU runtime finds the valid symbol and retrieves a function pointer dynamically. The user submits a string variable that matches the method's name to execute and the type-erased inputs and outputs required to execute the routine. Since the functions retrieved by \textit{dlsym} still have to be converted at compile-time to a function pointer with a static number of arguments (unlike FPGA and CUDA Driver API), the CPU runtime has to establish a limit to how many arguments a CPU-based submission can contain. Currently, the limit for CPU submissions is kernels with no more than ten arguments. Pointer arguments are implicitly converted to the correct type, and \textit{passing by value} required a \textit{reinterpret\_cast}. The thread pools are created by the runtime and default to the number of cores on the machines. 

The main pattern is a dynamic command processor with a task queue. Each submission generates an N-dimensional index table where the dimension is defined by the \textit{defer} or \textit{exec} methods. The indexing strategy is similar to existing work and is dictated by the parameters passed into the \textit{defer}, or \textit{exec} methods, in turn creating a Cartesian product of points and cycling through each combination during the execution of the kernels. The indices are generated pre-execution and are atomically or non-atomically accessed by the kernels via the kernel-side interface. Because atomic operations are expensive, additional space (padding) on the index tables is added for kernels whose correctness is not affected by repeated calls to a work item. If kernel correctness can be maintained, the CPU runtime can run in a non-atomic mode to improve performance for kernels with little work-per-work item combination. The kernel-side interface is quite simple and intuitive. The CPU runtime provides one kernel-side method to retrieve the N indices of the current work item. The method is a free function accessor taking the dimension of the index that needs to be returned as input. Since the CPU runtime (along with the other device runtimes) are singletons, a multi-threaded application will serialize the work making the free function thread-safe.

\begin{table*}[!hbt]
\scriptsize
\centering
\caption{\label{tab:benchmark} Comparison of host code control flow, SLOC, normalized framework overhead for two benchmark host codes implemented using different probability frameworks (for supporting Intel CPUs and NVIDIA GPUs, N=2).}
\begin{tabular}{|c|c|c|c|c|c|c|}
\hline
\textbf{HPC Kernels} & \textbf{Solutions} & \textbf{Device} & \textbf{\begin{tabular}[c]{@{}c@{}}Host Code Compile-time\\ Control Flow for supporting N \\Computing Devices\end{tabular}} & \textbf{SLOC} & \textbf{SLOC Reduction} & \textbf{\begin{tabular}[c]{@{}c@{}}Normalized Framework Overhead (\%)\\ (Normalized to Kernel Runtime)\end{tabular}} \\ \hline
\multirow{10}{*}{NBody} & \multirow{2}{*}{KOKKOS} & CPU & \multirow{2}{*}{N} & \multirow{2}{*}{15} & \multirow{2}{*}{1.5x} & 8.0e-5 \\ \cline{3-3} \cline{7-7} 
 &  & GPU &  &  &  & 7.2e-3 \\ \cline{2-7} 
 & \multirow{2}{*}{RAJA} & CPU & \multirow{2}{*}{N} & \multirow{2}{*}{18} & \multirow{2}{*}{1.8x} & 1.1e-3 \\ \cline{3-3} \cline{7-7} 
 &  & GPU &  &  &  & 3.2e-2 \\ \cline{2-7} 
 & \multirow{2}{*}{OpenMP} & CPU & \multirow{2}{*}{N} & \multirow{2}{*}{10} & \multirow{2}{*}{1x} & 24 \\ \cline{3-3} \cline{7-7} 
 &  & GPU &  &  &  & 4.4e-1 \\ \cline{2-7} 
 & \multirow{2}{*}{SyCL} & \multirow{2}{*}{CPU} & \multirow{2}{*}{N} & \multirow{2}{*}{21} & \multirow{2}{*}{2.1x} & \multirow{2}{*}{8.9} \\
 &  &  &  &  &  &  \\ \cline{2-7} 
 & \multirow{2}{*}{FLASH} & CPU & \multirow{2}{*}{1} & \multirow{2}{*}{10} & \multirow{2}{*}{-} & 9.8e-1 \\ \cline{3-3} \cline{7-7} 
 &  & GPU &  &  &  & 6.8 \\ \hline
\multirow{10}{*}{Particle-Diffusion} & \multirow{2}{*}{KOKKOS} & CPU & \multirow{2}{*}{N} & \multirow{2}{*}{12} & \multirow{2}{*}{1.2x} & 1.0e-5  \\ \cline{3-3} \cline{7-7} 
 &  & GPU &  &  &  & 1.2e-1 \\ \cline{2-7} 
 & \multirow{2}{*}{RAJA} & CPU & \multirow{2}{*}{N} & \multirow{2}{*}{17} & \multirow{2}{*}{1.7x} & 4.8e-3 \\ \cline{3-3} \cline{7-7} 
 &  & GPU &  &  &  & 5.2 \\ \cline{2-7} 
 & \multirow{2}{*}{OpenMP} & CPU & \multirow{2}{*}{N} & \multirow{2}{*}{10} & \multirow{2}{*}{1x} & 17 \\ \cline{3-3} \cline{7-7} 
 &  & GPU &  &  &  & 8.4 \\ \cline{2-7} 
 & \multirow{2}{*}{SyCL} & \multirow{2}{*}{CPU} & \multirow{2}{*}{N} & \multirow{2}{*}{22} & \multirow{2}{*}{2.2x} & \multirow{2}{*}{9.5} \\
 &  &  &  &  &  &  \\ \cline{2-7} 
 & \multirow{2}{*}{FLASH} & CPU & \multirow{2}{*}{1} & \multirow{2}{*}{10} & \multirow{2}{*}{-} & 1.9 \\ \cline{3-3} \cline{7-7} 
 &  & GPU &  &  &  & 13 \\ \hline
\end{tabular}
\end{table*}

\subsubsection{CUDA Runtime}
The CUDA runtime is effectively a wrapper around the CUDA Driver API. It supports a single-source and multi-source compilation process. For single-source cubins and/or PTX, pre-compiled code is extracted from the binary and is passed in through the \textit{RuntimeObj} through the implementation parameter. The CUDA runtime has similar logic as the CPU to convert human-readable symbols to their mangled counterpart. The kernel launches and launch parameters are forwarded to the \textit{cudaLaunch} method and run without change. The kernel implementations need to follow the normal CUDA toolchain compilation process, annotating functions with \textit{\_\_global\_\_} when appropriate. The CUDA runtime handles multiple devices and broker requests to different GPUs transparently. If the binary is built as a fat binary, the CUDA runtime will search for the appropriate binary for the launching device. This runtime will manage the data movement, synchronization, and devices for each job. The CUDA driver is much more flexible (see CUDA driver API vs. CUDA runtime API) for associating arguments to kernels. CUDA takes a double void star for encapsulating all their kernel arguments, and it is up to the higher-level interfaces and compiler to type-check declarations and convert accordingly.  

\subsubsection{OpenCL Runtime (for FPGAs) }

The OpenCL runtime is a wrapper around the Intel OpenCL runtime. This runtime manages multiple Intel FPGAs, along with the CUDA runtime, load-balances \textit{subactions} (kernels), and dispatches them uniformly in a round-robin manner. Kernel implementations are not available as single-source but are multi-source. For the best performance, we suggest early-loading the kernel files and creating the kernel object upon creating the \textit{RuntimeObj} to save time on loading and creating the \textit{clKernel} objects.  

\section{Evaluation}

We evaluate FLASH versus state-of-the-art portability frameworks and heterogeneous programming languages, including DPC++, KOKKOS, HIP, RAJA, and Legion, in terms of the complexity of framework usage (number of programming interfaces and objects) and the portability of the framework. Additionally, we statically analyzed an N-Body and particle-diffusion kernel across existing solutions and compared them to a FLASH implementation to assess the SLOC reduction and portability that can be achieved with FLASH. Furthermore, in order to avoid kernel and runtime implementation bias, we focus on the efficiency of the frameworks in terms of normalized framework overhead (normalized to the kernel runtime). This metric clearly explains the penalty each framework will induce on an application when invoked. Finally, we project the potential code portability improvement achieved by FLASH for a set of HPC applications and offer foresight on forward and backward compatibility on future heterogeneous platforms. 

\subsection{Experiment Setup}

\begin{table*}[]
\scriptsize
\centering
\caption{\label{tab:more_benchmark} Various refactored HPC Kernels (CUDA) host codes to FLASH. }
\begin{tabular}{|c|c|c|c|c|c|c|}
\hline
\textbf{HPC Kernels} & \textbf{Solutions} & \textbf{SLOC} & \textbf{SLOC Reduction} & \textbf{\begin{tabular}[c]{@{}c@{}}Normalized Framework Overhead (\%)\\ (Normalized to Kernel Runtime)\end{tabular}} \\ \hline
\multirow{2}{*}{1D Heat Transfer} & \multirow{1}{*}{CUDA}  & \multirow{1}{*}{12} & \multirow{1}{*}{4x} & \multirow{1}{*}{1.2e-6}  
\\ \cline{2-5}
& \multirow{1}{*}{FLASH} & \multirow{1}{*}{3} & \multirow{1}{*}{-} & 9
 \\ \hline 
\multirow{2}{*}{Attention} & \multirow{1}{*}{CUDA} & \multirow{1}{*}{22} & \multirow{1}{*}{3.6x} & 1.2e-6 
\\ \cline{2-5} 
& \multirow{1}{*}{FLASH} & \multirow{1}{*}{6} & \multirow{1}{*}{-} & 3.2
\\ \hline
\multirow{2}{*}{Black Scholes} & \multirow{1}{*}{CUDA} & \multirow{1}{*}{10} & \multirow{1}{*}{2.5x} & 1.5e-6 
\\ \cline{2-5} 
& \multirow{1}{*}{FLASH} & \multirow{1}{*}{4} & \multirow{1}{*}{-} & 5.2
\\ \hline
\multirow{2}{*}{Computational Fluid Dynamics} & \multirow{1}{*}{CUDA} & \multirow{1}{*}{54} & \multirow{1}{*}{4.15x} & 1.8e-6 
\\ \cline{2-5} 
& \multirow{1}{*}{FLASH} & \multirow{1}{*}{13} & \multirow{1}{*}{-} & 11
\\ \hline
\multirow{2}{*}{Convolution} & \multirow{1}{*}{CUDA} & \multirow{1}{*}{27} & \multirow{1}{*}{3x} & 1.6e-6 
\\ \cline{2-5} 
& \multirow{1}{*}{FLASH} & \multirow{1}{*}{9} & \multirow{1}{*}{-} & 2.0
\\ \hline
\multirow{2}{*}{Discrete Cosine Transform} & \multirow{1}{*}{CUDA} & \multirow{1}{*}{23} & \multirow{1}{*}{3.2x} & 1.2e-6
\\ \cline{2-5} 
& \multirow{1}{*}{FLASH} & \multirow{1}{*}{7} & \multirow{1}{*}{-} & 4
\\ \hline
\multirow{2}{*}{ISO2DFD} & \multirow{1}{*}{CUDA} & \multirow{1}{*}{22} & \multirow{1}{*}{3.14x} & 2.0e-6 
\\ \cline{2-5} 
& \multirow{1}{*}{FLASH} & \multirow{1}{*}{7} & \multirow{1}{*}{-} & 2.8
\\ \hline
\end{tabular}
\end{table*}

\raggedbottom
In order to compare the performance of FLASH with the existing portability framework, we measure the framework overhead and kernel runtime of each framework/kernel/Device tuple and calculate the normalized framework overhead by taking the ratio of framework overhead and kernel runtime. The framework overhead is the time needed for a blank kernel to be dispatched and completed. The kernel runtime is measured as the execution time taken by a kernel after dispatching and completion (not including data transfer to/from the host). This figure of merit represents the penalty imposed by a solution.

In order to measure the application code portability, we define Hardware Independence Ratio (HIR) as the following: 
\begin{align*}
    HIR = 1- Hardware\; Dependence\; Reduction\; (HDR). 
\end{align*}

To calculate HIR, we project the potential HDR achievable by FLASH. We statically analyzed a set of HPC applications and their hardware dependency on CUDA by downloading their latest main branches from the corresponding GitHub repositories. The statistics are approximated by running keyword searches on "CUDA", "OpenCL", "\_\_host\_\_", and "GPU". Files containing these keywords in object names or comments are counted as dependent. Only source files (in source folders) with well-known extensions (*.c, .f*, etc.) are analyzed. Makefiles, *.txt, .cmake, .md, data files or kernel files are not counted as source codes. 

Finally, the performance measurements in table \ref{tab:benchmark} were run on a dual-socket 10-core Intel 2650 v3 @ 2.3Ghz for all runs running on Ubuntu 18.04 LTS with 64GB 2133Mhz DRAM. The CPU-based performance measurement was done solely on the CPU cores with hyper-threading enabled. The GPU numbers were measured on the same host but with the main kernel offloaded to an NVIDIA Titan X running on CUDA toolkit 10.1. The code was compiled with GCC 10.1 for CPU runs and downgraded to GCC 4.8.5 for compatibility with NVCC (for host code).

\subsection{Experiment Results}

\begin{table}[tb]
\caption{\label{tab:AppTable} The potential code portability in terms of HIR achieved by FLASH calculated from the CUDA-dependency of different HPC application codes.}
\centering
\scriptsize
\begin{tabular}{|c|c|}
\hline
\textbf{Application} & \textbf{HIR of FLASH (\%)} \\ \hline \hline
GROMACS\cite{GROMACS_SRC}              & 90.1\%                   \\ \hline
NAMD\cite{NAMD_SRC}                 & 55.62\%                  \\ \hline
WRF\cite{WRF_SRC}                  & 99.72\%                   \\ \hline
GAMESS\cite{GAMESS_SRC}                & 86.95\%                  \\ \hline
LAMMPS\cite{LAMMPS_SRC}               & 92.11\%                   \\ \hline
CP2K\cite{CP2K_SRC}                 & 94.88\%                   \\ \hline
ParaView\cite{PARAVIEW_SRC}             & 98.54\%                   \\ \hline
MILC\cite{MILC_SRC}                 & 98.55\%                   \\ \hline
Chroma\cite{CHROMA_SRC}               & 99.27\%                   \\ \hline
AutoDock - GPU\cite{AUTODOCK_SRC}       & 54.85\%                  \\ \hline
Abinit\cite{ABINIT_SRC}               & 94.71\%                   \\ \hline
specfem3d\cite{SPECFEM_SRC}            & 94.84\%                   \\ \hline
\end{tabular}
\end{table}
\raggedbottom

Table \ref{tab:benchmark} shows the comparison of host code compile-time control flow, SLOC, and normalized framework overhead for two benchmark host codes (N-Body and Particle Diffusion) implemented using four existing solutions (KOKKOS, RAJA, OpenMP, SyCL) and FLASH on CPU and GPU. The FLASH SLOC is based on the host code available in the FLASH repository. Table \ref{tab:benchmark} compares the SLOC of the host codes as they pertain to the preparation of the runtime (\textit{ie KOKKOS::Intialize}), device (\textit{i.e., device\_selector}), and inputs (\textit{i.e., buffers}), as well as dispatching tasks (\textit{i.e., RAJA::ForAll, submits, parallel\_for}). The SLOC does not include kernel definitions since they vary highly depending on developer experience and algorithm.  The comparison results show a reduction of SLOC that ranges from 1x to 2.2x, primarily resulting from the preparation of the runtime, device, and, most importantly, input and dispatch (lambda functions). The performance varies based on a combination of framework and kernel implementation. However, all simulations maintain the same parameters across frameworks, such as the number of particles and iterations. OpenMP notably has the most overhead for both kernels on CPU at 24\% and 17\%, respectively. FLASH has less than 2\% overhead on CPU, with 6\% and 13\% on the GPU. This is directly attributed to inefficiencies in the CUDA device runtime logic and will be optimized in later releases. FLASH alleviates data synchronization and preparation across single and multiple submissions by managing data preparation to the device runtime. The overhead across frameworks was nearly all inconsequential and served as a fraction of the runtime, with OpenMP taking the most notable time vs. the kernel runtime at a fourth of the runtime for OMP target offload. Table \ref{tab:more_benchmark} shows additional benchmarks with their host codes refactored from CUDA to FLASH. The kernels are still written in CUDA; however, their host codes have been rewritten to be dispatched from the FLASH runtime (using the CUDA backend). This table demonstrates the completeness of the API to support more complex codes with minimal framework overhead. The higher framework overhead numbers come from the embedded host and device side loop controls and predicate logic. The SLOC count reduction demonstrates the flexibility of the host code to defer dispatching the to frontend FLASH runtime, allowing for greater interoperability/portability from the host application. Furthermore, although this table only shows the combination of FLASH with the CUDA backend, with more time, it can be easily demonstrated for multiple runtimes (and devices) without host code modifications.  Overall, FLASH can achieve up to 4.0x SLOC reduction, especially for the benchmarks with high preparation costs.

Table \ref{tab:AppTable} shows the code portability achieved by FLASH measured by HIR. The percentage HIR values are projected using the possible reduction in the hardware dependency of CUDA of various HPC application codes. Based on the portability exhibited by the available FLASH host code templates, we project that one can completely decouple the existing accelerator dependency from the host code and kernel, resulting in an achievement of up to 99.72\% application code portability. This allows the applications to be more easily refactored and accelerator-agnostic for future hardware platforms. 

\section{Conclusion}

This paper presented FLASH 1.0, a C++-based software framework for rapid parallel deployment and enhancing host code portability in heterogeneous computing. 
Making the programming interfaces completely vendor-agnostic, hardware-agnostic, and task-agnostic is critical to enabling a single control flow for supporting various heterogeneous accelerators as well as a lean library with a small, limited number of programming interfaces and objects that is sustainable as the supported accelerator devices keep growing. Such unification of control flow is the key to enabling the true portability of host codes, and such simplicity of the framework is the key to easy usage and adoption.  

\bibliographystyle{IEEEtran}
\bibliography{flash_workcited.bib}

\vspace{12pt}
\end{document}